**Deep Learning-Based Grading of Ductal Carcinoma *In Situ* in Breast Histopathology Images**

Suzanne C. Wetstein[1,#], Nikolas Stathonikos[2], Josien P.W. Pluim[1], Yujing J. Heng[3], Natalie D. ter Hoeve[2], Celien P.H. Vreuls[2], Paul J. van Diest[2] and Mitko Veta[1]

1. Medical Image Analysis Group, Department of Biomedical Engineering, Eindhoven University of Technology, Eindhoven, The Netherlands
2. Department of Pathology, University Medical Center Utrecht, University Utrecht, Utrecht, The Netherlands
3. Department of Pathology, Harvard Medical School, Beth Israel Deaconess Medical Center, Boston, MA, USA

[#]Corresponding Author: Suzanne Wetstein, Medical Image Analysis Group, Department of Biomedical Engineering, Groene Loper 5, 5612AE Eindhoven, The Netherlands; Email: s.c.wetstein@tue.nl; Phone: +31 40-247 5581

**Abstract**

Ductal carcinoma *in situ* (DCIS) is a non-invasive breast cancer that can progress into invasive ductal carcinoma (IDC). Studies suggest DCIS is often overtreated since a considerable part of DCIS lesions may never progress into IDC. Lower grade lesions have a lower progression speed and risk, possibly allowing treatment de-escalation. However, studies show significant inter-observer variation in DCIS grading. Automated image analysis may provide an objective solution to address high subjectivity of DCIS grading by pathologists.

In this study, we developed and evaluated a deep learning-based DCIS grading system. The system was developed using the consensus DCIS grade of three expert observers on a dataset of 1186 DCIS lesions from 59 patients. The inter-observer agreement, measured by quadratic weighted Cohen's kappa, was used to evaluate the system and compare its performance to that of expert observers. We present an analysis of the lesion-level and patient-level inter-observer agreement on an independent test set of 1001 lesions from 50 patients.

The deep learning system (dl) achieved on average slightly higher inter-observer agreement to the three observers (o1, o2 and o3) ($\kappa_{o1,dl}$ = 0.81, $\kappa_{o2,dl}$ = 0.53 and $\kappa_{o3,dl}$ = 0.40) than the observers amongst each other ($\kappa_{o1,o2}$ = 0.58, $\kappa_{o1,o3}$ = 0.50 and $\kappa_{o2,o3}$ = 0.42) at the lesion-level. At the patient-level, the deep learning system achieved similar agreement to the observers ($\kappa_{o1,dl}$ = 0.77, $\kappa_{o2,dl}$ = 0.75 and $\kappa_{o3,dl}$ = 0.70) as the observers amongst each other ($\kappa_{o1,o2}$ = 0.77, $\kappa_{o1,o3}$ = 0.75 and $\kappa_{o2,o3}$ = 0.72). The deep learning system better reflected the grading spectrum of DCIS than two of the observers.

In conclusion, we developed a deep learning-based DCIS grading system that achieved a performance similar to expert observers. To the best of our knowledge, this is the first automated system for the grading of DCIS that could assist pathologists by providing robust and reproducible second opinions on DCIS grade.

**Introduction**

Breast cancer remains one of the leading causes of death in women [1]. Most breast cancers are invasive ductal carcinomas (IDCs) which arise from epithelial cells lining the ducts. Ductal carcinoma *in situ* (DCIS) refers to the pre-invasive stage whereby the cancer cells remain contained within the basement membrane. Studies suggest that a considerable part of DCIS lesions may never progress into IDC [2-4]. Autopsy studies indicate that occult DCIS exists in 9% (range 0–15%) of women [5]. A few small-scale studies have been done on patients where misdiagnosis of DCIS led to the omission of surgery. In the course of 30 years, 14-53% of these patients developed IDC [6-8]. A meta-analysis of multiple studies of patients with DCIS showed a 15-year total (invasive and noninvasive) local recurrence rate of 40% after a diagnosis of DCIS on excisional biopsy [9]. Invasive cancer was found in 28% of recurrence cases, with an 18% cancer-specific mortality rate [9]. Another study found that the 20-year breast cancer-specific mortality rate following a diagnosis of DCIS was 3.3% [10]. Thus, there is a substantial portion of DCIS lesions that may never develop into IDC.

Since it is challenging to predict which patients will and will not progress to IDC [11], the diagnosis of DCIS prompts immediate surgical treatment. This decision is currently made regardless of the histologic grade of the DCIS lesion, while lower grade lesions have a lower progression speed and risk [12]. High-grade DCIS cases represent 42-53% of total cases [13-16] and are considered to have a high risk for recurrence [15, 17-19] and breast cancer-specific mortality [10].

In view of the perceived need to de-escalate treatment of DCIS, ongoing clinical trials (LORD [2], LORIS [3], COMET [20, 21] and LARRIKIN [22]) aim to monitor disease progression of patients with low risk DCIS (based on the histologic grade of their DCIS lesions) that forgo surgical treatment. As such, accurate histologic grading of DCIS is crucial for the clinical management of these patients. The aforementioned clinical trials utilize different classification systems to grade DCIS lesions as good (grade 1), moderate (grade 2) or poor (grade 3) differentiation. Histologic grading systems commonly used in practice are the Van Nuys classification [23], Holland classification [24], and Lagios classification [25]. Schuh *et al*. [26] compared grading among 13 pathologists using these three grading systems and 43 DCIS cases. They found that all systems had at best moderate agreement, the best being the Van Nuys classification (κ=0.37). Other studies also showed significant inter-observer variation in DCIS grading, regardless of the grading system used [27-32].

Methods that have been studied to improve inter-observer agreement in DCIS grading include evaluating B7-H3 expression [33], using laboratory- and pathologist-specific feedback reports and e-training [34, 35] and applying ad hoc dichotomization of histopathologic features to obtain 'ideal' cut-offs for features used in DCIS grading [36]. Although these studies showed promise in reducing grading variation, overall grading variation remained substantial.

Dano *et al.* [36] recommended the use of machine learning to help tackle this diagnostic challenge. Indeed, the subjectivity and low reproducibility of histologic DCIS grading make it amenable for automated assessment by image analysis. Automated systems have the potential to decrease the workload of pathologists and standardize clinical practice [37, 38]. Deep learning based grading and survival prediction have been previously applied to histopathology images [38-42] and deep neural network models have been successfully developed for other tasks specific to breast histopathology [43-49]. State-of-the-art deep convolutional neural networks (CNN) have been shown to outperform pathologists in detecting metastases in sentinel lymph nodes of breast cancer patients [50]. Previous work combining machine learning and DCIS was done by Bejnordi *et al.* [51] who designed a successful DCIS detection algorithm that works fully automatically on whole slide images (WSIs). The system detects epithelial regions in WSIs and classifies them as DCIS or benign/normal (i.e. malignant tissue was not a part of this study). Eighty percent of DCIS lesions were detected, at an average of 2.0 false positives per WSI. This system only detects DCIS lesions and does not grade them.

In this paper we describe the development of an automated deep learning DCIS grading system. Our system is the first of its kind. It was developed using consensus grades based on grading by three expert observers and also incorporates the uncertainty in DCIS grading between these expert observers. In an independent test set, we compared the DCIS grading results at both lesion- and patient-level between our deep learning system and three expert observers.

**Materials and Methods**

*Study design and population*

Digital slides were retrieved from the digital pathology archive of the University Medical Center in Utrecht, The Netherlands, from cases dated between Jan 1, 2016 and Dec 31, 2017, for patients who underwent a breast biopsy or excision and were labeled with 'ductal carcinoma *in situ*'. This included all cases that contained DCIS regardless of the main diagnosis (i.e. cases with IDC and DCIS were also included). Since images were used anonymously, informed consent was not needed. For each patient up to three representative hematoxylin and eosin (H&E) stained WSIs containing DCIS lesions were selected by expert observers. In total, 116 WSIs from 109 patients were included in this study. The slides were scanned using the Nanozoomer 2.0-XR (Hamamatsu Phonics Europe GmbH, CJ Almere, The Netherlands) at ×40 magnification with a resolution of 0.22 μm per pixel.

*Pathological assessment*

Histologic grading into grades 1, 2 or 3 was performed according to the Holland classification system [24]. This classification system is recommended by The Netherlands Comprehensive Cancer Organisation [52] and focuses on nuclear morphologic and architectural features. Low grade nuclei have a monotonous appearance and a small size not much larger than normal epithelial cell size. Nucleoli and mitoses only occur occasionally. In contrast, high grade nuclei show marked pleomorphism, are large in size and contain one or more conspicuous nucleoli. Intermediate grade nuclei are defined as neither low nor high grade [53]. Architecturally, low grade DCIS is cribriform and/or micropapillary, while high grade DCIS is solid and often shows central necrosis.

All DCIS lesions present in the 116 WSIs were annotated by two experienced pathologists and one pathology assistant who grades cases on a regular basis. This was done using the open-source software Automated Slide Analysis Platform (ASAP; Computation Pathology Group, Radboud University Medical Center, Nijmegen, The Netherlands). Each lesion was outlined by one observer and independently graded by all three observers. In total, 2187 lesions were annotated.

A consensus grade for each DCIS lesion was obtained by majority voting. In the case where all three observers gave a different grade the assigned consensus grade was grade 2. For the expert observers, the DCIS grade at the patient-level was assigned as the highest lesion grade present for the respective patient, although patients can have heterogeneous lesions [54].

*Development of the deep learning system*

For the development and validation of the deep learning system the 109 patients were randomly assigned to three distinct subsets: training, validation (used for model selection and parameter tuning) and test datasets, whilst ensuring the distribution of DCIS grades was similar in each subset. The training set contained 879 DCIS lesions from 40 patients, the validation set contained 307 lesions from 19 patients and the test set contained 1001 lesions from 50 patients.

The data acquisition process resulted in WSI with outlined DCIS lesions. The DCIS lesions were extracted from the WSI by fitting a rectangular box around the manually annotated lesions. An additional 90μm border was drawn around these boxes in order to include the DCIS lesion as well as the surrounding stroma. The stroma was included because tumor-associated stroma has been shown to be detected in greater amounts around DCIS grade 3 than DCIS grade 1 [46] and DCIS associated stromal changes might play a role in progression to IDC [55]. The boxes were extracted at magnification level ×10.

The deep learning system developed to grade DCIS takes into account the inter-observer variability in DCIS grading. The system was trained on two targets: 1) the consensus of the DCIS grades given by the three expert observers, and 2) the number of observers that agreed with this consensus grade. This was done as we believe there can be extra information in the inter-observer variability in DCIS annotations. A lesion that was annotated as grade 1 by two observers and as grade 2 by the third observer is probably a borderline case, while a lesion that was annotated as grade 1 by all three observers is more clear-cut. By giving the system the information which cases in the training set are borderline cases it might be able to learn the distinction between the grades better. To evaluate the added value of the inclusion of observer agreement we compared our deep learning system with a baseline system which was trained on the consensus DCIS grades only. Our deep learning system outperformed this baseline system on the validation set.

The deep learning system was based on the Densenet-121 [56] network architecture. As input to the network we cropped a random patch of 512 × 512 pixels (about 450 μm × 450 μm) from a DCIS lesion and used data augmentation to overcome the variability of the tissue staining appearance, which is an important hurdle in histopathology image analysis [57]. The deep learning system was trained on patches extracted from the training dataset. The validation dataset was used to monitor the performance of the network during training and to prevent overfitting. All further results shown in this paper will be results on the independent test set. During evaluation on the test dataset, we extracted 10 randomly located patches from one lesion and took the median of the predicted grades as the predicted grade. No data augmentation was used during test time. More details on the deep learning system and its hyper-parameters can be found in Supplementary Information. The deep learning system developed in this study will be made available for scientific and non-commercial use through our Github page. The dataset will also be made publicly available via the grand-challenge.org platform.

As stated before, for the expert observers the DCIS grade at the patient-level was determined by the highest lesion grade present for the respective patient. Expert observers grade a case by examining all DCIS lesions in a WSI while the algorithm is only shown one lesion at a time. Examining all lesions at once might lead to grading lesions more similarly, leading to "regression to the mean". To mimic the practice of expert observers, we chose to let the automated patient-level grade be determined by the lesion at the $P^{th}$ percentile, where the value of $P$ was determined by best patient-level DCIS grading performance on the validation dataset. For the deep learning system, this resulted in the patient-level grade being determined by the lesion at the 80$^{th}$ percentile.

*Statistical analysis*

The inter-observer and model-vs-observer agreement for the DCIS grading was measured using quadratic weighted Cohen's Kappa. This measure is commonly used for inter-rater agreement on an ordinal scale because it compensates for the degree of error in category assessment. This means that disagreement by one grade point is weighted less than disagreement by two grade points. Using this method, each observer was compared with every other and Kappa values were recorded for each pairing. All analyses were performed using Python version 3.6 and the deep learning model was implemented using the Keras deep learning framework [58].

**Results**

*Population characteristics*

Patient- and lesion-level characteristics for our test dataset are summarized in Table 1. Mean patient age was 58 years (95% CI: 55 – 61 years) and the number of lesions per patient was 20 (95% CI: 11 – 30). Using the consensus grade of the three observers, there were seven patients with DCIS grade 1, 24 patients with grade 2 and 19 patients with grade 3. The average lesion area was 0.48mm$^2$ (95% CI: 0.37 – 0.60 mm$^2$). There were 152 grade 1 lesions, 645 grade 2 lesions and 204 grade 3 lesions.

**Table 1:** Patient and lesion characteristics in the test dataset.

| Patient characteristics | |
|---|---|
| *n* | 50 |
| Age at biopsy | |
|     Mean years (95% CI) | 58 (55 – 61) |
| Number of lesions per patient | |
|     Mean *n* (95% CI) | 20 (11 – 30) |
| Consensus grade (*n*, %) | |
|     Grade 1 | 7 (14%) |
|     Grade 2 | 24 (48%) |
|     Grade 3 | 19 (38%) |
| Lesion characteristics | |
| *n* | 1001 |
| Lesion size | |
|     Mean mm$^2$ (95% CI) | 0.48 (0.37 – 0.60) |
| Consensus grade (*n*, %) | |
|     Grade 1 | 152 (15%) |
|     Grade 2 | 645 (64%) |
|     Grade 3 | 204 (20%) |

*Lesion-level inter-observer agreement in test dataset*

Inter-observer agreement on DCIS grading at the lesion-level between three expert observers and the deep learning system is shown in Table 2. Inter-observer agreement between expert observers was κ = 0.58, κ = 0.50 and κ = 0.42. The deep learning system showed agreement with the observers of κ = 0.81, κ = 0.53 and κ = 0.40. The average agreement between expert observers was lower than that between expert observers and the deep learning system. The confusion matrices of inter-observer agreement on DCIS grading between expert observers and the deep learning system are shown in Figure 1. Interestingly, there was high agreement between observers 2 and 3 for grade 2 (563 lesions), but both observers graded more than half of the lesions as grade 2 (observer 2: 693 out of 1001, observer 3: 749 out of 1001 lesions). In contrast, observer 1 and the deep learning system graded only 481 and 512 lesions, respectively, as grade 2.

**Table 2:** Inter-observer quadratic weighted Cohen's Kappa for ductal carcinoma in situ (DCIS) grading at the lesion-level among three observers and the deep learning system. The results are shown on the test set which contains 1001 lesions from 50 different patients. The 95% confidence interval (CI) was determined analytically.

|  | Expert observers | | Deep learning system |
|---|---|---|---|
|  | Observer 2 | Observer 3 |  |
|  | κ (95% CI) | κ (95% CI) | κ (95% CI) |
| Observer 1 | 0.58 (0.49 – 0.66) | 0.50 (0.41 – 0.60) | 0.81 (0.75 – 0.86) |
| Observer 2 | - | 0.42 (0.29 – 0.54) | 0.53 (0.44 – 0.62) |
| Observer 3 | - | - | 0.40 (0.29 – 0.54) |

**Figure 1:** Confusion matrices for DCIS grading at the lesion-level between observers **(A)** and between observers and the deep learning system **(B)**. These are results on the test set which contained 1001 DCIS lesions from 50 patients.

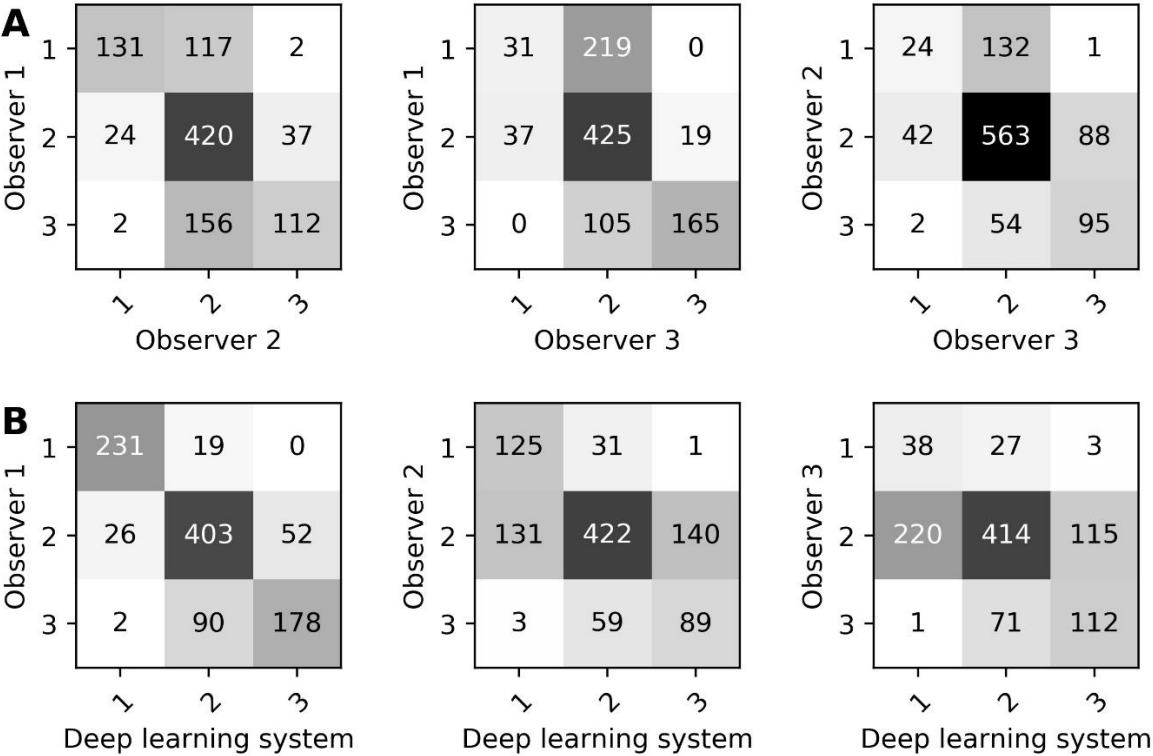

Ten lesions from five patients had high disagreement (i.e., cases being assigned grade 1 and grade 3) between two expert observers or between an expert observer and the deep learning system (Figure 2). These lesions were all very small with an area ≤ 0.06 mm², while the average lesion area was 0.48 mm² (95% CI: 0.37 – 0.60 mm²). Upon review of the lesions, by a consensus meeting of the three expert observers, two lesions concerned small isolated detachments (floaters) and five were potentially incorrectly annotated as DCIS. For the remaining three lesions, the deep learning system classified two correctly (the high disagreement was caused by one of the expert observers) and one incorrectly.

**Figure 2:** All lesions with high disagreement between expert observers and between expert observers and the deep learning system. Lesions with the same letter come from the same patient. All lesions had an image size of 512 × 512 pixels except for **(B2)** where we show the middle 512 × 512 pixel patch. For lesion **(A)** the observers graded 2-3-2 and the deep learning system predicted grade 1. On final review in a consensus meeting, grades 1 and 3 did not seem justified, therefore the expert observers assigned this lesion as grade 2. For lesions **(B1, B2)** the observers graded 1-3-1 and the deep learning system predicted grade 1. Grade 3 did not seem justified during the consensus meeting and was an error by an expert observer. For lesion **(C1)** the observers graded 3-2-2 and the deep learning system predicted grade 1. For lesion **(C2)** the observers graded 3-2-3 and the deep learning system predicted grade 1. Both these lesions concern floaters and should not have been in the dataset. For lesion **(D1)** the observers graded 3-1-2 and the deep learning system predicted grade 2. For lesion **(D2)** the observers graded 3-1-3 and the deep learning system predicted grade 2. On review, both lesions are not obviously DCIS. For lesion **(E1)** the observers graded 2-1-1 and the deep learning system predicted grade 3. For lesions **(E2, E3)** the observers graded 2-2-1 and the deep learning system predicted grade 3. On review, these three lesions are not obviously DCIS.

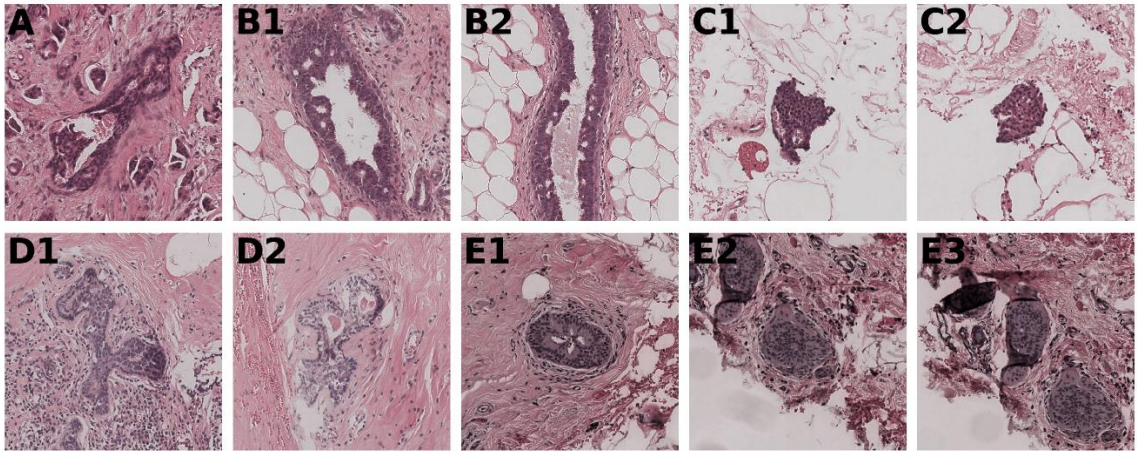

*Patient-level inter-observer agreement in test dataset*

Inter-observer agreement between three expert observers and the deep learning system on DCIS grading at the patient-level is shown in Table 3. Inter-observer agreement between expert observers was κ = 0.77, κ = 0.75 and κ = 0.72. The deep learning system showed agreement with the observers of κ = 0.77, κ = 0.75 and κ = 0.70. The average agreement between expert observers was slightly higher than that between expert observers and the deep learning system. The confusion matrices for DCIS grading at the patient-level by expert observers and our deep learning system are shown in Figure 3. At the patient level, there was no large disagreement (i.e. no case assigned as grade 1 and grade 3) between two expert observers or between an expert observer and the deep learning system.

**Table 3:** Inter-observer quadratic weighted Cohen's Kappa for ductal carcinoma in situ (DCIS) grading at the patient-level amongst three observers and the deep learning system. The results are shown on the test set which contains 50 patients. The 95% confidence interval (CI) was determined analytically.

|  | Expert observers | | Deep learning system |
|---|---|---|---|
|  | Observer 2 | Observer 3 |  |
|  | κ (95% CI) | κ (95% CI) | κ (95% CI) |
| Observer 1 | 0.77 (0.49 – 1.05) | 0.72 (0.40 – 1.04) | 0.77 (0.49 – 1.05) |
| Observer 2 | - | 0.75 (0.46 – 1.04) | 0.70 (0.41 – 1.00) |
| Observer 3 | - | - | 0.75 (0.46 – 1.04) |

**Figure 3:** Confusion matrices for DCIS grading at the patient-level between observers **(A)** and between observers and the deep learning system **(B)**. These are results on the test set which contained 50 patients.

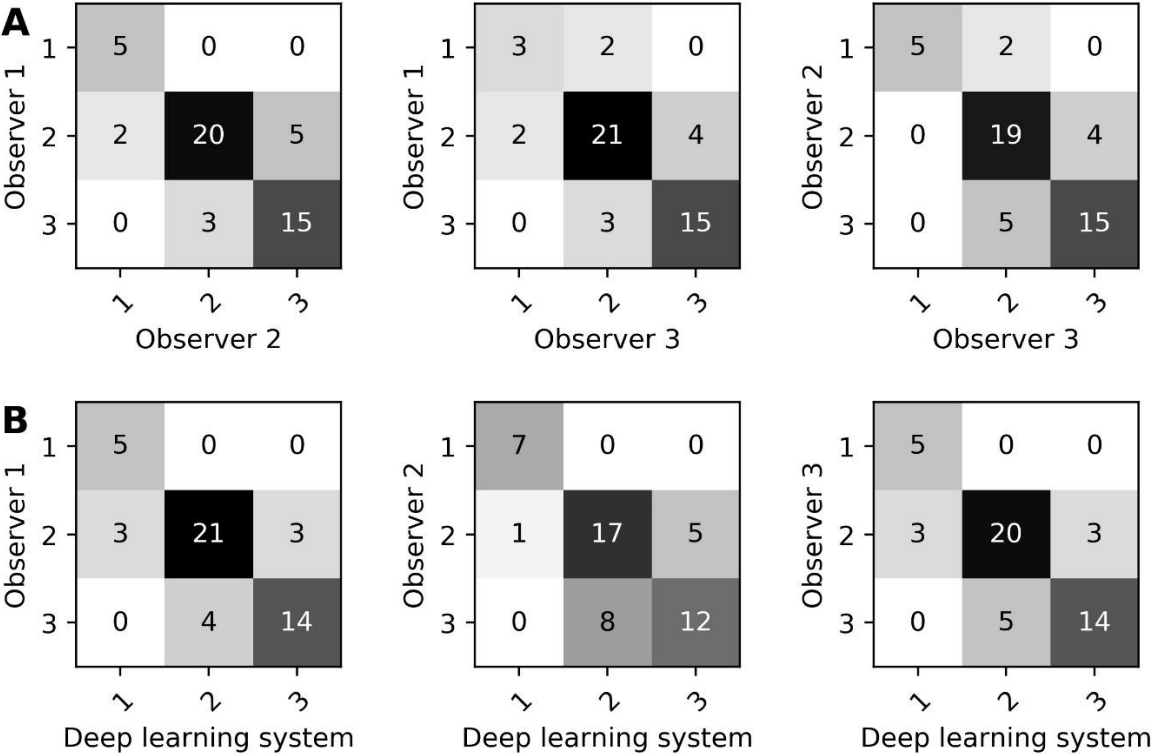

**Discussion**

DCIS currently prompts immediate surgical treatment while a considerable part of DCIS lesions may never progress into IDC. Lower grade lesions progress into IDC less often and at a slower pace. However, grading systems dividing DCIS lesions into low/middle/high grade were shown to have significant inter-observer variation. Accurate and reproducible DCIS grading may be possible with the help of automated image analysis.

We developed a fully automated deep learning system to grade DCIS lesions of the breast. To the best of our knowledge, this is the first automated classification system for the grading of DCIS. Our study demonstrated that our automated system achieved a performance similar to expert observers. The system has the potential to improve DCIS grading by acting as a reliable and consistent first or second reader.

In this study, three observers achieved an average quadratic weighted Kappa score at the lesion-level of 0.50. This is similar to Douglas-Jones *et al.* [30] who found an average quadratic weighted Kappa score of 0.48 between 19 pathologists for Van Nuys classification of 60 DCIS lesions. Other studies have also shown high inter-observer variability for DCIS grading but used other measures of agreement, like unweighted Cohen's Kappa or percentage of agreement between observers [26-29, 31, 32]. The high inter-observer variability between observers reiterates the need to create a consistent system for DCIS grading.

We developed a deep learning system to grade DCIS that incorporates the inter-observer variability (aleatoric uncertainty) in the training process. At the lesion-level, the system achieved agreement with observers comparable to observers amongst each other. The confusion matrices (Figure 1) showed high agreement between observer 1 and the deep learning system. These matrices also showed that observers 2 and 3 graded many cases (69% and 75%, respectively) as grade 2. In contrast, observer 1 and the deep learning system graded only 48% and 51% of lesions, respectively, as grade 2. Therefore, grading by both observer 1 and the deep learning system was more diversified, better reflecting the grading spectrum of DCIS.

We found high disagreement (i.e., cases being assigned grade 1 and grade 3) between observers and between observers and the deep learning system in ten lesions as shown in Figure 2. Re-examination of these lesions showed that two lesions concerned small isolated detachments (floaters) and five were potentially incorrectly annotated as DCIS. Since these errors only occurred in 7 out of 1001 lesions (i.e., 0.7%), we decided neither to exclude the floaters and wrongly annotated lesions from the dataset nor rerun the analysis as it would unlikely yield significantly better results. For the remaining three lesions, the deep learning system classified two correctly (one of the expert observers caused the high disagreement) and one incorrectly. All of these lesions were very small (area ≤ 0.06 mm$^2$). We hypothesize that both expert observers and deep learning have difficulties classifying tiny lesions because there is less information to work with, and observers may have graded these lesions not so much on their specific morphologic appearance but similar to the larger surrounding ones, leading to "regression to the mean".

At the patient-level, the expert observers were slightly more in agreement with each other than with the deep learning system. The confusion matrices (Figure 3) show that there were no grade 1 vs. grade 3 discrepancies between expert observers nor between expert observers and the deep learning system. Expert observers amongst each other agreed more on grade 3, whereas expert observers and the deep learning system agreed more on grade 1.

The main limitation of this study is that the dataset used to develop our deep learning system originated from a single medical center. Although we applied data augmentation to expand our training dataset, the robustness of the system can be improved by including WSIs from different institutions obtained using different whole slide scanners and different staining protocols.

Our system was able to achieve a performance similar to that of expert observers. Due to the fact that the system was trained on data annotated by expert observers it would be hard to exceed their performance. For future studies it would be interesting to gather information on whether patients with DCIS progressed to IDC. This information could be used to train a deep learning system to predict which DCIS lesions have a high chance of progressing to IDC. Realistically, it would only be possible to gather follow up information for patients with low grade DCIS as it would not be safe to forgo treatment for patients with high grade DCIS. Data could possibly be gathered from low grade DCIS patients that entered clinical trials (e.g. LORD [2], LORIS [3], COMET [20, 21] and LARRIKIN [22]). The information on which low grade DCIS lesions progress to IDC within 5-10 years could be used to improve the grading practice of pathologists and automated systems.

In conclusion, we developed and evaluated an automated deep learning-based DCIS grading system which achieved a performance similar to expert observers. With further evaluation, this system could assist pathologists by providing robust and reproducible second opinions on DCIS grade.

**Acknowledgements**

This work was supported by the Deep Learning for Medical Image Analysis research program by The Dutch Research Council P15-26 and Philips Research (SCW, MV and JPWP).

**Conflict of Interest**

The authors declare no conflict of interest.

**Supplementary Information**

*Input to the deep learning system*

The data acquisition process resulted in whole-slide images (WSI) with outlined ductal carcinoma *in situ* (DCIS) lesions. The DCIS lesions were extracted from the WSI by fitting a rectangular box around the manually annotated lesions. An additional 90μm border was drawn around these boxes in order to include the DCIS lesion as well as the surrounding stroma. These boxes were extracted at magnification level ×10. We cropped random patches of 512 × 512 pixels (about 450 μm × 450 μm) from all DCIS lesion boxes and applied data augmentation during training (not at test time). The data augmentation consisted of translations, rotations, flipping, shearing, zooming and color channel shifts. All augmentations were applied randomly within certain limits. Translations of up to 25% of the image size were applied horizontally and vertically. Images were rotated up to ±90 degrees and flipped horizontally and vertically. The shear intensity, range for zoom and color channel intensity were all changed up to ±20% of their original values. After data augmentation the patches were fed into the network.

*The deep learning system*

The deep learning system used to grade DCIS was based on the DenseNet-121 [1] network architecture. The mini batch size was set to 12 and we balanced the batches to always include 4 patches of each DCIS grade. The networks were trained by minimizing the ordinal categorical cross-entropy loss between the ground truth and the predictions. This loss was calculated as the categorical cross-entropy loss multiplied with the differences in the predicted and ground truth DCIS grade plus 1. We used ImageNet pre-training and the optimization was done with stochastic gradient descent with a learning rate of 1e-4 and momentum of 0.95. Training was stopped when the kappa score on the validation set started decreasing. Hyper-parameters, like the learning rate and the mini batch size, were tuned to optimal performance on the validation set by grid search.

*Output of the deep learning system*

The first output of the system was a DCIS grade (either 1, 2 or 3) for each patch. The second output was a number that predicted how many observers would agree with this grade (either 1, 2 or 3). We did not use this second prediction at test time, but it was used during training to feed the network information about boundary cases.

*Variations across training runs of the deep learning system*

Deep learning methods can have different predictive performance across training runs because of the stochastic nature of gradient descent and the fact that different local minima may be found for each run. To research the variability in DCIS grading for our deep learning system we trained it 5 times with the same hyper-parameter settings. The results in the main manuscript are the results from the first training run. Table 1 shows the mean and standard deviation of the inter-observer agreement over 5 training runs.

**Table 1:** Inter-observer quadratic weighted Cohen's Kappa for ductal carcinoma in situ (DCIS) grading at the lesion-level and patient-level between three observers and the deep learning system. The results are shown on the test set which contains 1001 lesions from 50 different patients. The Cohen's Kappa shown is the mean of 5 training runs of the same deep learning algorithm. The standard deviation between the 5 training runs is also shown.

|  | Lesion-level | | Patient-level | |
| --- | --- | --- | --- | --- |
|  | Deep learning system | | Deep learning system | |
|  | mean($\kappa$) | SD($\kappa$) | mean($\kappa$) | SD($\kappa$) |
| Observer 1 | 0.76 | 0.03 | 0.69 | *0.06* |
| Observer 2 | 0.50 | 0.02 | 0.70 | 0.04 |
| Observer 3 | 0.41 | 0.01 | 0.74 | 0.03 |